\documentclass[12Pt]{article} 
\usepackage{amssymb,euscript,latexsym,amsmath} 
\usepackage{graphicx} 
\usepackage{epstopdf}
\usepackage{bm} 
\usepackage{enumerate} 
\usepackage{color} 
\usepackage{setspace} 
\usepackage{subfigure} 
\usepackage{hyperref}

\begin{document}

\title{Effects of Body Elasticity on Stability of  Underwater Locomotion} 
\author{Fangxu Jing and Eva Kanso}

\maketitle 
\begin{abstract}
We examine the stability of  the ``coast" motion of fish, that is to say, the motion of a neutrally buoyant fish at constant speed in a straight line. The forces and moments acting on the fish body are thus perfectly balanced. The fish motion is said to be unstable if a perturbation in the conditions surrounding the fish results in forces and moments that tend to increase the perturbation and it is stable if these emerging forces tend to reduce the perturbation and return the fish to its original state. Stability may be achieved actively or passively. Active stabilization requires neurological control that activates musculo-skeletal components  to compensate for the external perturbations acting against stability.  Passive stabilization on the other hand requires no energy input by the fish and is dependent upon the fish morphology, i.e. geometry and elastic properties.
In this paper, we use a deformable body consisting of an articulated body equipped with torsional springs at its hinge joints and submerged in
an unbounded perfect fluid as a simple model 
to study passive stability as a function of the body geometry and spring stiffness. We show that for given body dimensions, the spring elasticity, 
when properly chosen, leads to passive stabilization of the (otherwise unstable) coast motion.

\end{abstract}

\section{Introduction}
\label{sec:intro}

We analyze the passive stability of a class of elastic bodies moving in a perfect fluid. 
Our primary motivation is to examine the role of body elasticity as a mechanism for passively 
stabilizing the coast motion of fish. 
Fish seem to alternate between actively swimming (``burst") and passively 
responding (``coast") to the surrounding fluid -- a cycle known as the ``burst and coast" cycle, see~\cite{ViWe1982}. 
It is of course difficult to gauge whether the coast motion is entirely passive, that is, whether or not the fish employs
active control for stabilization purposes; see, for example,~\cite{Weihs2002}. Yet, experimental evidence
seems to suggest the existence of some mechanisms for passive stabilization in fish motion. For example,
\cite{BeHoTrLiLa2006} reported upstream motion of anesthetized fish (undeniably passive) 
in vortical flows and mentioned that the elastic properties of the fish body are essential to achieve this motion. 
Indeed, the motion of elongated \textit{rigid} bodies in unsteady flows is known to be \textit{passively unstable}, 
see, for example,~\cite{Leonard1997, Fish2002, OsKa2011}. 
Body elasticity may not be the only mechanism responsible for passively stabilizing such  motion.
Other physical mechanisms such as the effects of the fish wake or the boundary layer 
effects may also be at play. In this paper, we focus solely on the role of body elasticity in the passive stability of 
the coast motion. This is of course a mathematical idealization while, in reality,  the effects of body elasticity, 
viscosity and vortex shedding are all coupled and operate simultaneously.
This mathematical idealization is aimed to untangle the effects of body elasticity on the stability
of motion and is viewed as a first step in constructing models with higher degrees of complexity and fidelity to the physical system.

%

\begin{figure}
\begin{center}
\subfigure[unstable for all $b/a<1$]{\includegraphics[scale=0.55]{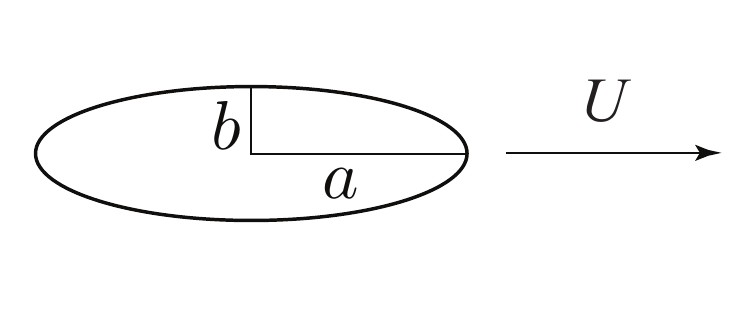}}	\hspace{0.25in}
\subfigure[passively stable for a range of $b/a<1$ and $k$ values]{\includegraphics[scale=0.55]{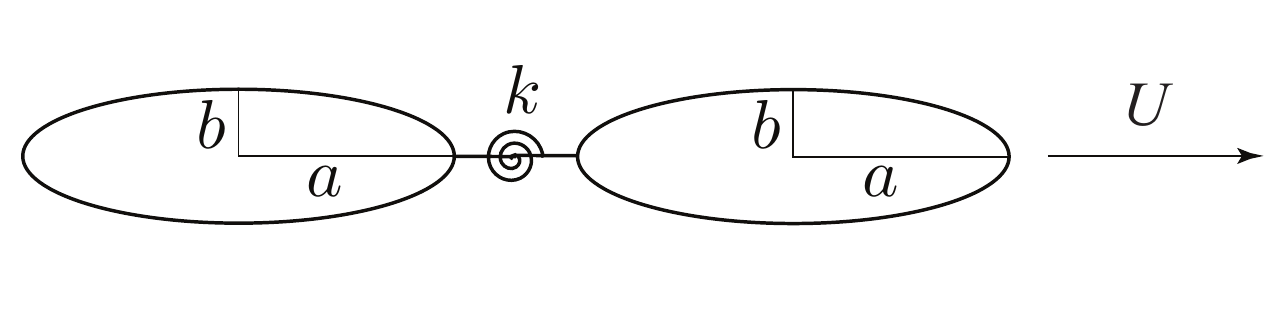}} \vspace{0.1in}
\subfigure[passively stable for a range of $b/a<1$ and $k_{1}, k_2$ values]{\includegraphics[scale=0.55]{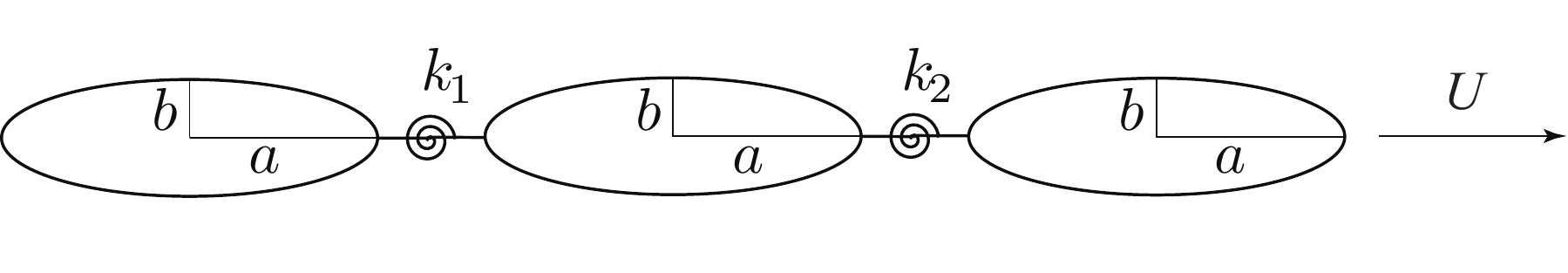}} \vspace{0.1in}
\end{center}
\caption{\footnotesize  Family of relative equilibria for articulated body moving in an inviscid, incompressible fluid: (a) rigid ellipse is unstable for $b< a$; (b) and (c) two-link and three-link bodies may be stable for proper choice of spring stiffness versus body dimensions.} 
\label{fig:equilibria} 
\end{figure}

We model the fish as an articulated body formed of $(n+1)$ rigid links, say, identical ellipses 
of major- and minor-axes $a$ and $b$, respectively. The links are connected via
$n$ hinge joints with torsional springs at the joints (with stiffness $k_{i}, i = 1, ..., n$) to emulate the elasticity of the fish body. 
The articulated body is submerged in an unbounded volume of incompressible, inviscid 
fluid at rest at infinity. 
The dynamics of this body-fluid system admits a family of relative equilibrium solutions
where the body, in its elongated configuration, is steadily moving along its major axis of 
symmetry, see figure~\ref{fig:equilibria}. We refer to these translational motions as the
coast motion.
We examine the effects of the springs' elasticity on the stability of the coast motion. 
We focus on passive stability, that is to say, the stability of motion when the deformable 
body is subject to initial conditions only and is passively responding to the 
surrounding fluid. This is in contrast to a deformable body  that is actively controlling its shape to swim (see~\cite{KaMaRoMe2005}) or to stabilize its motion (see~\cite{Weihs2002}). While the motion of a single
rigid body, see figure~\ref{fig:equilibria}$(a)$  is known to be passively unstable, we find that, for the elastic body, there exists a range of parameter
values -- spring stiffness versus body geometry -- for which the coast motion is passively stable. We examine the regions of stability
for a two-link and a three-link body. We find that the stable region of the two-link body is characterized by a bending mode of deformations.
For the three-link body, there exists two stable regions, one is characterized by a bending mode of deformation and the second by a 
traveling-wave mode of deformation.


This problem is somewhat reminiscent to that of the fluttering flag instability. 
The latter has been the focus of a large number of experimental (\cite{ShVaZh2005}), computational (\cite{Alben2008,MiSmGl2008}), 
and theoretical (\cite{ArMa2005}) investigations. Of course, the flag is held stationary at one end
in an incoming uniform flow field whereas the fish is freely moving.
This constraint does alter the forces applied on the flag and, in turn, may significantly
influence the stability results. For example,~\cite{ShMaBuKe2002} and~\cite{KaOs2008} showed that the stability of the freely moving F\"{o}ppl equilibria
is distinct from the stability of the stationary equilibria in an incoming uniform flow field. Further, the flag is assumed to be heavy whereas the fish is
neutrally-buoyant. The hydrodynamic forces acting on the flag are thus dominated by vortex shedding rather than the added mass effect. 
All is to say that, while the fish and flag are somewhat conjugates, 
the fluttering flag instability (typically analyzed as a function of the flag's rigidity versus weight)
cannot be readily translated to the  stability of the moving fish which we analyze in this 
work as a function of the body stiffness versus geometry.


\section{Problem Setting}

\begin{figure}
	\begin{center}
	\includegraphics[scale=0.5]{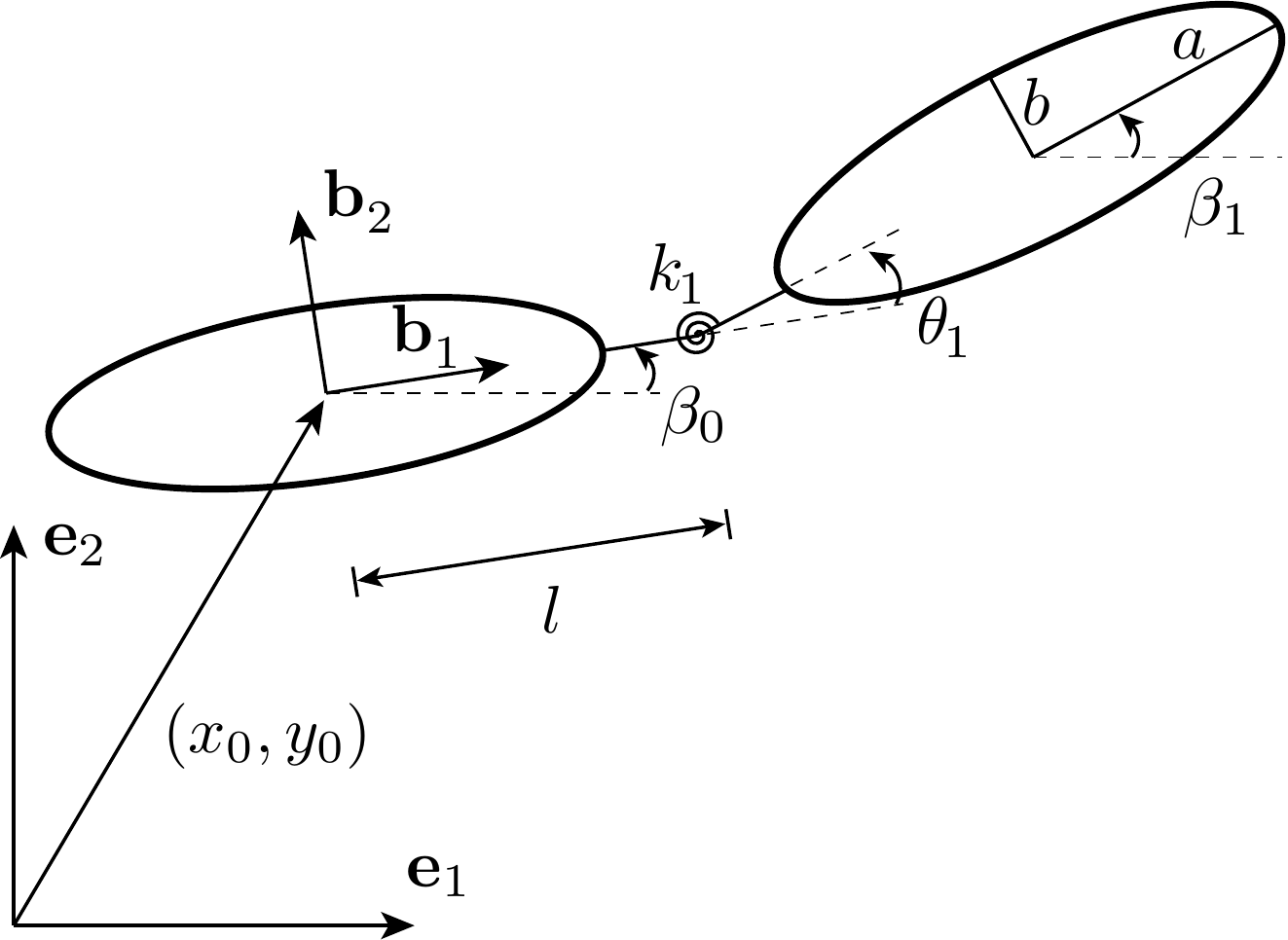}
	\caption{\footnotesize Schematic of a two-link body.}
		 \label{fig:model}
	\end{center}
\end{figure}

Consider an articulated body moving in an unbounded volume of incompressible, inviscid 
fluid at rest at infinity.  The articulated body is composed of 
$n+1$ identical ellipses (with semi-axes $a$ and $b$) of uniform density $\rho$
equal to that of the fluid. That is to say, the articulated body is neutrally-buoyant. 
The ellipses or links forming the body are connected via $n$ massless 
and frictionless hinge joints placed at a distance $l$ from the ellipses' centers 
along the major axes and equipped with torsional springs of 
constant stiffness $k_{i}$, $i=1,\ldots, n$. 

Let $\{\mathbf{e}_{1,2,3}\}$ be an orthonormal inertial frame where $\{\mathbf{e}_{1,2}\}$
spans the plane of motion. The configuration of the articulated body can then be described 
by $2(n+1)$ position coordinates $(x_j,y_j)$ of the ellipses' centers
and $(n+1)$ orientation variables  $\beta_j$ (where $j=0,\ldots,n$) defined as the angles between the ellipses' major axes of symmetry and the $\mathbf{e}_1$-axis,
see figure~\ref{fig:model}.
This set of $3(n+1)$ coordinates is subject to $2n$ holonomic constraints due to the presence of the hinge joints
connecting the ellipses.
These constraints are implicitly accounted for when describing the body's configuration in terms of $(x_0, y_0)$,  
$\beta_0$, and $n$ relative angles $\theta_i=\beta_{i+1}-\beta_{i}$ ( $i=1,\ldots,n$) that denote the angles between 
every two consecutive links of the body. The coordinates $(x_0,y_0,\beta_0)$ coincide with the inertial coordinates of the first 
ellipse and are called locomotion coordinates while the relative angles $\theta_i$ describe only the shape of the articulated body.
For notational convenience, we introduce the $(n+3)$-dimensional configuration vector
$\mathbf{q} \equiv\{x_0 , y_0 , \beta_0, \theta_1, \ldots, \theta_{n}\}^T$, where $()^T$ denotes the transpose
and the $n$-dimensional shape vector $\bm{\Theta}\equiv \{\theta_1, \ldots, \theta_{n}\}^T$.

It is convenient for deriving the equations of motion in \S\ref{sec:eom} to write the velocity of the articulated body 
in terms of an orthonormal body-fixed frame $\{\mathbf{b}_{1,2,3}\}$ (with $\mathbf{b}_3 \equiv \mathbf{e}_3$)
whose origin is attached at $(x_0,y_0)$ and such that 
$\beta_0$ is the angle between $\mathbf{b}_1$ and $\mathbf{e}_1$. 
Let $(u_j,v_j)$ denote the components of the translational velocity vector $\mathbf{v}_j$ ($j=0,\ldots, n$)
of the center of each ellipse expressed in the body frame $\{\mathbf{b}_{1,2,3}\}$. In other words, let $\mathbf{v}_j = u_j \mathbf{b}_1 + v_j \mathbf{b}_2$. 
Also, let  $\bm{\Omega}_j = \Omega_j \mathbf{b}_3$ 
denote the angular velocity vector of each ellipse.  One has $\Omega_0 = \dot{\beta}_0$ and $ \Omega_j=\dot{\beta}_0 + \sum_{i=1}^j \dot{\theta}_i$ for $j = 1,\ldots, n$. 
Here, the dot notation $\dot{()}$ is used to denote derivative with respect to time $t$. 
We now construct a $3(n+1)$-dimensional velocity vector $\bm{\xi} 
\equiv\{u_0 , v_0 , \Omega_0, u_1 , v_1 , \Omega_1, \ldots, u_n , v_n , \Omega_n\}^T$.
The velocity vector $\bm{\xi}$ can be related to the velocity
vector $\dot{\mathbf{q}}$ via a  $3(n+1)\times(3+n)$ transformation matrix $\mathbb{R}$, function of $\beta_0$ and $\bm{\Theta}$ only, namely,
\begin{equation}\label{eq:constraint} 
\bm{\xi} = \mathbb{R} \dot{\mathbf{q}}.
\end{equation}
For example, in the case of an articulated body made of two ellipses connected via one hinge joint, the $6\times4$ transformation matrix $\mathbb{R}$ is
\begin{equation}
\begin{split}
\mathbb{R} = \left( \begin{array}{cccc} 
\cos \beta_0 & \sin\beta_0 & 0 & 0 \\
-\sin \beta_0 & \cos\beta_0 & 0 & 0 \\
0 & 0 & 1 & 0 \\
\cos \beta_0 & \sin\beta_0 & -l \sin \theta_1 & -l\sin \theta_1 \\
-\sin \beta_0 & \cos\beta_0 & l (1 + \cos \theta_1) & l \cos \theta_1\\
0 & 0 & 1 & 1 \\
\end{array} \right).
\end{split}
\end{equation}
This transformation matrix can be readily generalized to arbitrary but finite $n$.

%
%
%
%


\section{Equations of motion}
\label{sec:eom}

We derive the equations of motion for the fluid-mass system using Hamilton's 
least action principle for which the Lagrangian function is given by
\begin{equation}
\mathcal{L}= T - V
\end{equation}
where  $T=T_{B}+T_{F}\,$ denotes the kinetic energy of the fluid-mass system, whereas $V$ denotes the potential energy stored in the springs
and can be written in matrix form as
\begin{equation}
V={1\over 2}\bm{\Theta}^T\mathbb{K}\bm{\Theta}\,,
\label{equation:U}
\end{equation}
where the  stiffness matrix $\mathbb{K}$ is an $n\times n$ diagonal matrix with entries $k_i$ corresponding to the stiffness of the torsional springs at the hinge joints.

The kinetic energy $T_{B}$ of the articulated body is given by
%
\begin{equation}
T_B = {1\over 2}\bm{\xi}^T  \mathbb{M}_{\rm body}\bm{\xi} \, , 
\label{eq:TB}
\end{equation}
where $ \mathbb{M}_{\rm body}$ is a $3(n+1) \times 3(n+1)$ diagonal mass matrix with $3\times3$ block diagonal entries of the form
 \begin{equation}
\left( \begin{array}{ccc}
m_{\rm ellipse}&0 &0  \\
0&m_{\rm ellipse}&0  \\
0&0& J_{\rm ellipse}\\
\end{array} \right),
\end{equation}
where $m_{\rm ellipse} = \rho\pi ab$ and $J_{\rm ellipse} = \rho\pi ab (a^2+b^2)/4$ are the actual mass and moment of inertia of each ellipse.

In potential flow,  the kinetic energy of the fluid $T_{F}$
can be written, using standard vector identities and techniques (see for example~\cite{KaMaRoMe2005} and~\cite{NaKa2007}), 
as a function of the configuration and velocity of the submerged body,
\begin{equation}
T_F = {1\over 2}\bm{\xi}^T  {\mathbb{M}}_{\rm added} \bm{\xi} \, .
\label{eq:TF}
\end{equation}
The added mass matrix ${\mathbb{M}}_{\rm added}$ is a $3(n+1)\times 3(n+1)$ symmetric
matrix that accounts for the presence of the fluid. Differently said, in potential flow,
the hydrodynamic forces and moments acting on the surfaces of the accelerating bodies can be
completely accounted for by `adding mass'  to  the submerged bodies (see~\cite{Br1982}
for a comprehensive review). The entries of the added mass matrix $\mathbb{M}_{\rm added}$
depend on the shape of the articulated body which may be changing in time as dictated by the 
evolution of the shape variable $\bm{\Theta}$.
We compute the entries of ${\mathbb{M}}_{\rm added}(\bm{\Theta})$ numerically using a panel method, as discussed
in~\cite{KaMaRoMe2005,Kanso2009}. It is worth noting that, although the added mass effect captures all
the hydrodynamic forces in the potential flow environment, this is not the case in real viscous fluids where one needs
to account for additional drag forces and forces induced by vortex shedding and vortex wakes as discussed in~\S\ref{sec:discussion}.

Adding~\eqref{eq:TB} and~\eqref{eq:TF} and substituting~\eqref{eq:constraint} 
into the resulting expression, the total kinetic energy $T=T_B + T_F$ of the fluid-mass system can be written as 
\begin{equation}
T = {1\over 2}\dot{\mathbf{q}}^T \mathbb{M}\dot{\mathbf{q}}\, ,
\qquad \mathbb{M}\ = 
\mathbb{R}^T\left( \mathbb{M}_{\rm body} + \mathbb{M}_{\rm added}(\bm{\Theta}) \right)
\mathbb{R}  \, ,
\label{eq:T_total}
\end{equation}
where $\mathbb{M}$
is an $(n+3)\times(n+3)$ mass matrix as a function of $\beta_0$ and $\bm{\Theta}$ only since the transformation matrix
$\mathbb{R}$ depends only on $\beta_0$ and $\bm{\Theta}$.
The Lagrangian function can then be expressed as
\begin{equation}
\mathcal{L}(\mathbf{q},\dot{\mathbf{q}})=\dfrac{1}{2} \dot{\mathbf{q}}^T \mathbb{M} \dot{\mathbf{q} }-
{1\over 2} {\bm{\Theta}}^T \mathbb{K}{\bm{\Theta}}\, .
\label{eq:lagrangian}
\end{equation}
In the absence of external forces and moments acting on the fluid-body system, the equations of motion
 are given by the $(n+3)$ Euler-Lagrange equations,
\begin{equation}
\dfrac{\text{d}}{\text{d}t}(\dfrac{\partial \mathcal{L}}{\partial \dot{\mathbf{q}}}) - 
\dfrac{\partial \mathcal{L}}{\partial \mathbf{q}} = 0 \, . 
\label{eq:eom_2ndorder}
\end{equation}
This class of problems is energy-preserving, hence the total energy $E = \dfrac{1}{2} \dot{\mathbf{q}}^T \mathbb{M} \dot{\mathbf{q} }
+ \dfrac{1}{2} \bm{\Theta}^T \mathbb{K} \bm{\Theta}$ is conserved.
It also admits two integrals of motion since the entries of the $(n+3)\times(n+3)$ mass matrix $\mathbb{M}$
depend exclusively on $\beta_0$ and $\bm{\Theta}$
and not on the position $(x_0,y_0)$ of the articulated body.
The coordinates $x_0\,,y_0$ are thus ignorable coordinates of the Lagrangian $\mathcal{L}$ and the associated linear momenta 
$p_x = \partial{\mathcal{L}}/\partial{\dot{x}_0}$ and $p_y = \partial{\mathcal{L}}/\partial{\dot{y}_0}$ are conserved. These two symmetries reflect the 
fact that the system's dynamics is invariant to rigid translations of the whole body. Additionally, the system possesses a rotational symmetry or invariance to
rigid rotations of the whole body. This rotational symmetry is associated with conservation of the 
system's total angular momentum $h = \partial \mathcal{L}/\partial \dot{\beta}_0 + {x}_0 (\partial \mathcal{L}/\partial \dot{y}_0) - {y}_0 (\partial \mathcal{L}/\partial \dot{x}_0)$.
The conservation of linear and angular momenta $p_x$, $p_y$ and $h$ can be used to eliminate the variables $x_0$, $y_0$ and $\beta_0$ from the equations of motion.
That is to say, the $(n+3)$ second-order differential equations of motion obtained from~\eqref{eq:eom_2ndorder} can be rewritten, by virtue of these three conserved quantities and a standard procedure, as $(2n+3)$ first-order differential equations 
governing the body-frame velocities $u_0(t)$, $v_0(t)$, $\Omega_0(t)$, the shape variables $\bm{\Theta}(t)$ and their time derivatives $\dot{\bm{\Theta}}(t)$. 
Namely, one can rewrite~\eqref{eq:eom_2ndorder} in the form
\begin{equation}
	\dot{\boldsymbol{\eta}} = \mathbf{f}(\boldsymbol{\eta})\,,\label{eq:eom}
\end{equation}
where we introduced the compact notation
\begin{equation}
\boldsymbol{\eta} = \{u_0 \,, v_0 \,, \Omega_0 \, , \bm{\Theta}, \dot{\bm{\Theta}}\}^T\,.
\end{equation}
The function $\mathbf{f}(\boldsymbol{\eta})$ is a nonlinear $(2n+3)$ vector-valued function of $\bm{\eta}$ whose explicit form is omitted here for brevity. 
To get an idea of the form of these equations, note that when the articulated body is formed of a single ellipse with no deformation variables, that is to say,
the body is rigid, \eqref{eq:eom} reduce to three first-order differential equations in $u_0,v_0,\Omega_0$ which are exactly the Kirchhoff's equations 
of motion for a submerged elliptic body; see, for example,~\cite{Lamb1932,Leonard1997}.
For an articulated body with $n$ joints, the three `Kirchhoff-like' equations are coupled to $2n$ first-order 
differential equations governing the shape deformations $\bm{\Theta}(t)$ of the body, $n$ of these equations
are simply stating that the time derivate of $\bm{\Theta}$ is equal to $\dot{\bm{\Theta}}$.

Equation~\eqref{eq:eom} admits a family of relative equilibrium solutions of the form ${\boldsymbol{\eta}}_e = [\,U \,, \, 0 \,, \, 0 \,,\, 0\,, \ldots, 0 \,]^T$, 
where $U$ is an arbitrary constant. This relative equilibrium corresponds to the motion that the articulated body in its elongated configuration when 
all its links are translating along their semi-major axis at an arbitrary coast speed $U$, with zero rotation and zero relative angles. We refer to this family 
of equilibria as the coast motion because it reminds of the coast phase in the fish burst and coast cycle. Our goal in the next section is to understand the 
role of elasticity in the passive stability of these equilibria.

\section{Stability Analysis}
\label{sec:stability}

We use linearization techniques to examine the stability of the equilibrium ${\boldsymbol{\eta}}_e = [\,U \,, \, 0 \,, \, 0 \,, 0\, , \ldots\,, 0 \,]^T$ to the initial perturbations. 
We begin by presenting numerical evidence, via numerically integrating the nonlinear equations~\eqref{eq:eom}, that the system is indeed passively
stable for certain parameter values. 

Before we proceed, we non-dimensionalize the equations of motion~\eqref{eq:eom} by scaling the length with $a$ (the length of the semi-major axis of the ellipses) 
and the time with $a/U$ whereas the mass is scaled with $\rho\pi a^2$. All the parameters and variables can then be written in non-dimensional form accordingly. For example, one gets
\begin{equation}
\begin{split}
	\tilde{a} & = \frac{a}{a} = 1,\quad \tilde{b} = \frac{b}{a},\quad \tilde{U} = \frac{U}{U} = 1,\quad \tilde{k}_i = \frac{k_i}{\rho\pi a^2 U^2},\\[2ex]
	\tilde{m}_{\rm ellipse} & = \frac{\rho\pi ab}{\rho\pi a^2}=\tilde{b},\quad \tilde{J}_{\rm ellipse} = \frac{\rho\pi ab (a^2+b^2)/4}{\rho \pi a^4}=\dfrac{\tilde{b} (1+\tilde{b}^2)}{4},.
\end{split}
\end{equation}
Entries in the added mass matrix can be made non-dimensional in a similar way. In this study, we assume $\tilde{l} = 1.1$. Note that, after this scaling, the number of independent parameters
reduces to the geometric parameter $\tilde{b}$ and the spring stiffness $\tilde{k}_i$. We want to understand the passive stability of the relative equilibrium
$\tilde{\bm{\eta}}_e = [\,1 \,, \, 0 \,, \, 0 \,, 0\, , \ldots\,, 0 \,]^T$
as a function of the dimensionless parameters $\tilde{b}$ and $\tilde{k}_i$.
For convenience, we drop the tilde~$\tilde{()}$~ notation with the understanding that all variables are non-dimensional hereafter.



\begin{figure}[!t]
	\begin{center}	
	\subfigure[$k = 0.2$]{
	\includegraphics[scale=0.28]{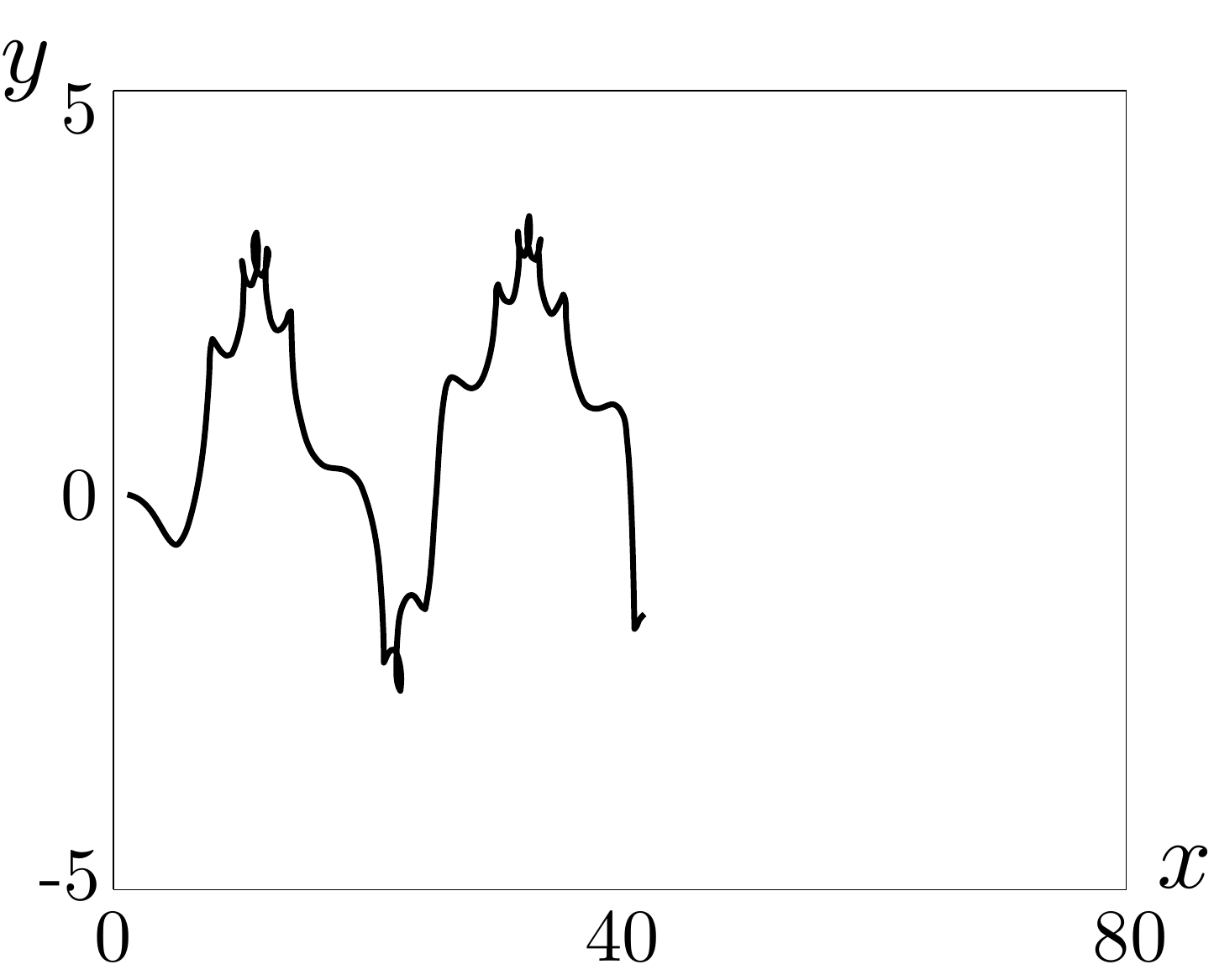} }
	\subfigure[$k = 0.35$]{
	\includegraphics[scale=0.28]{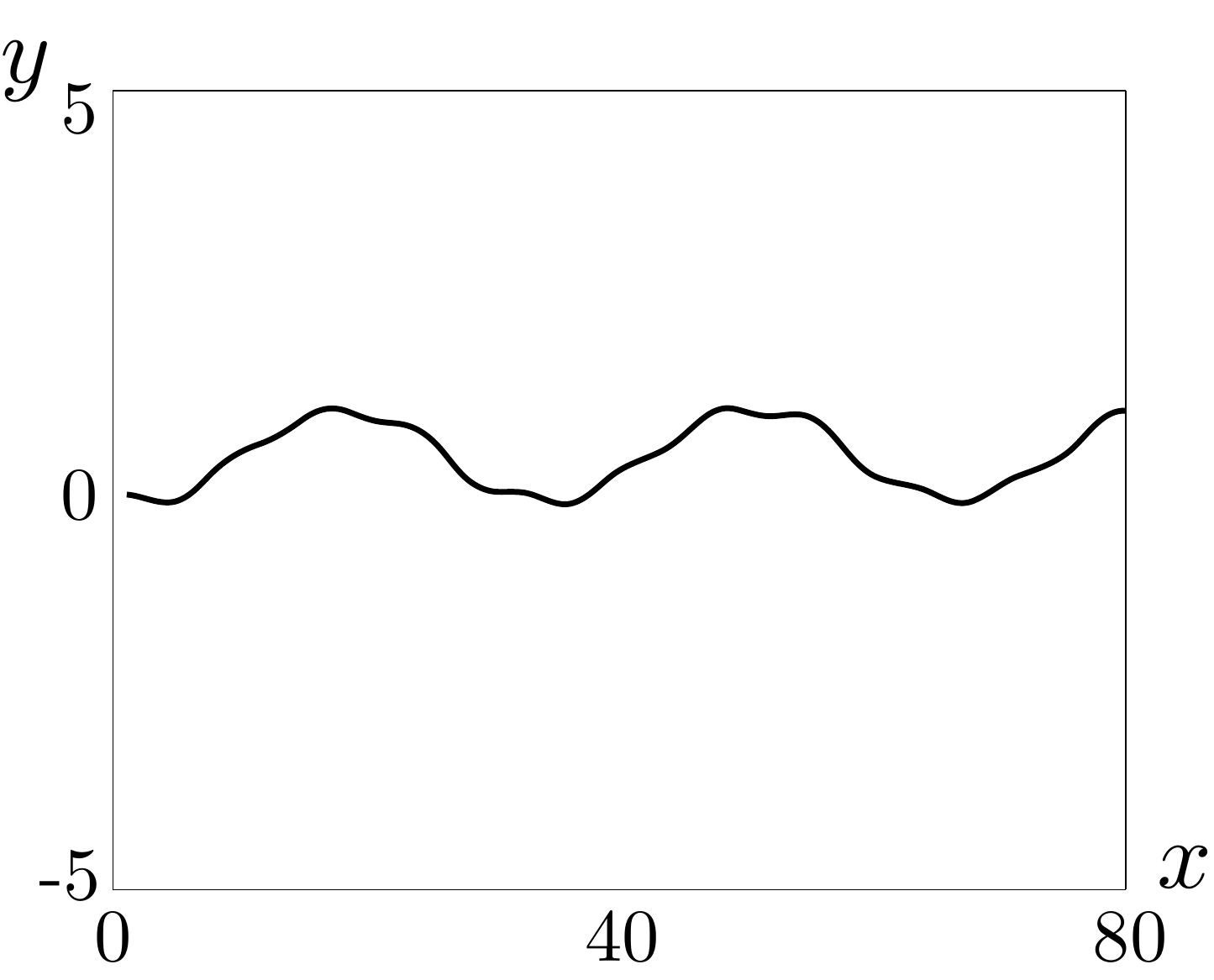} }
	\subfigure[$k = 0.5$]{
	\includegraphics[scale=0.28]{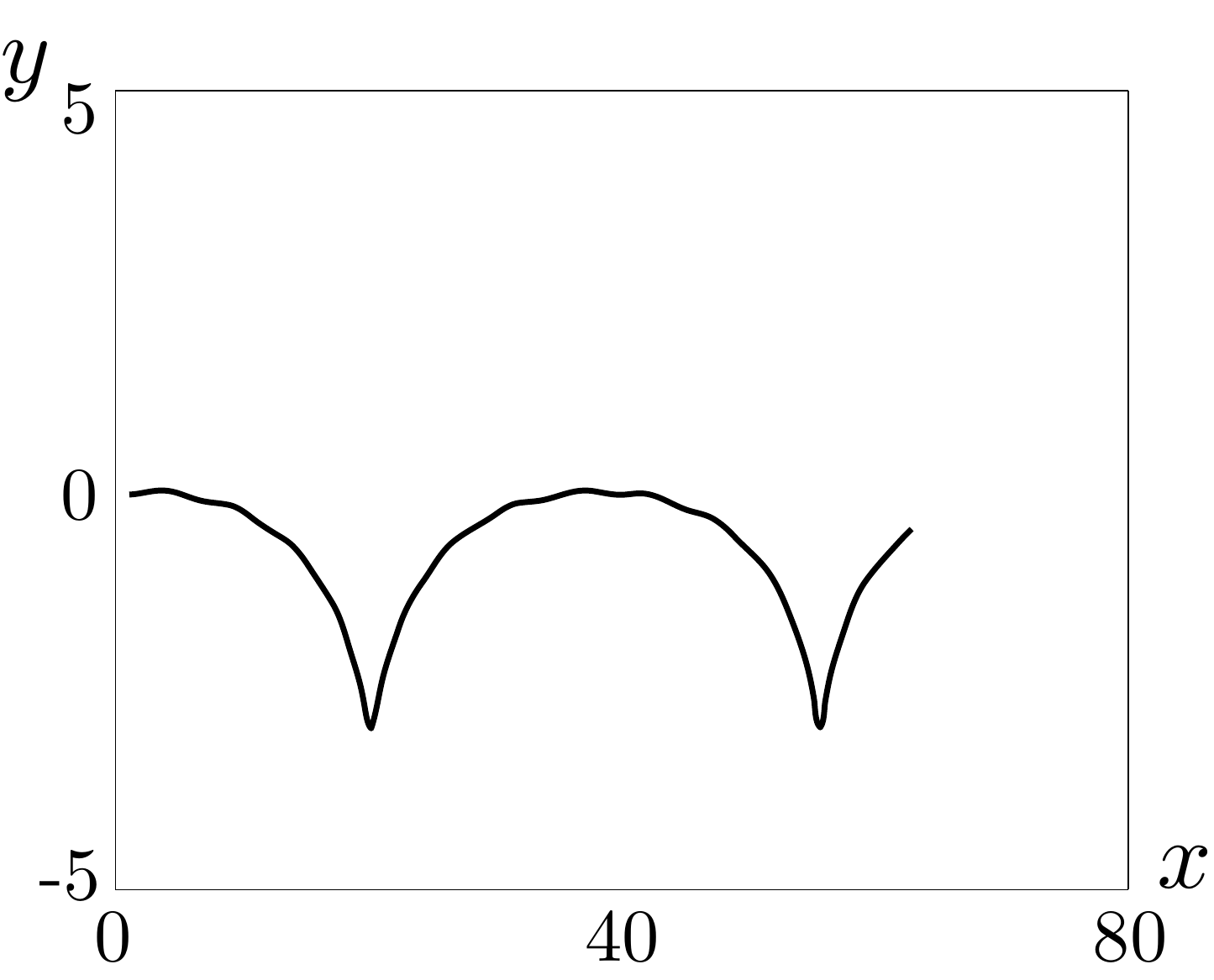} }
		\caption{\footnotesize Planar $(x_0,y_0)$ trajectories of two-link fish for $b=0.2$ and (a) $k=0.2$, (b) $k=0.35$ and (c) $k=0.5$.   Initial condition is $\boldsymbol{\eta}_0 = [\, 1\,, 0\,, 0\,, 0.1\,, 0]^T$. Total integration time is 80.} \label{fig:trajectory}
	\end{center}
\end{figure}
\begin{figure}[!t]
	\begin{center}	
	\subfigure[$k = 0.2$]{
	\includegraphics[scale=0.28]{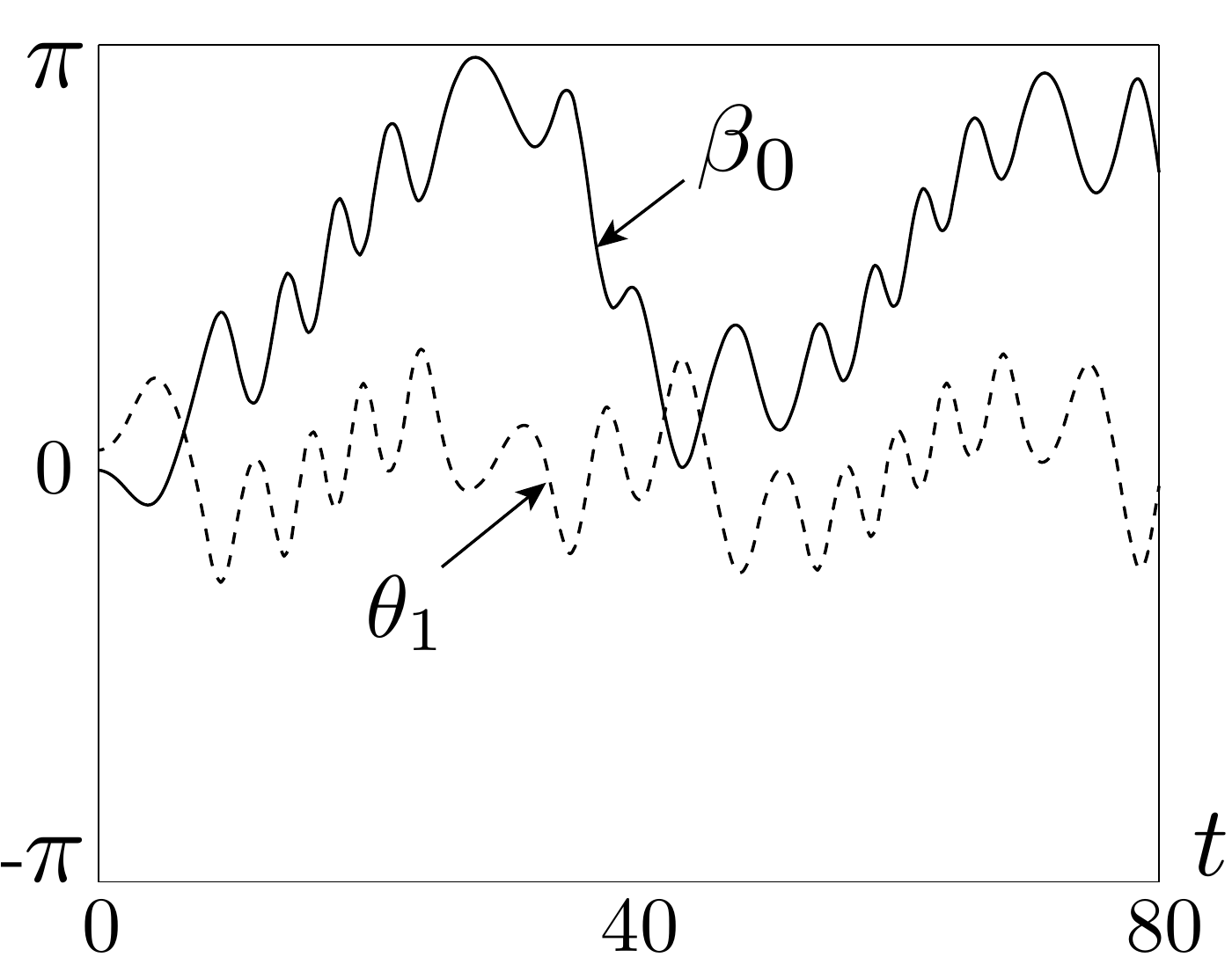} }
	\subfigure[$k = 0.35$]{
	\includegraphics[scale=0.28]{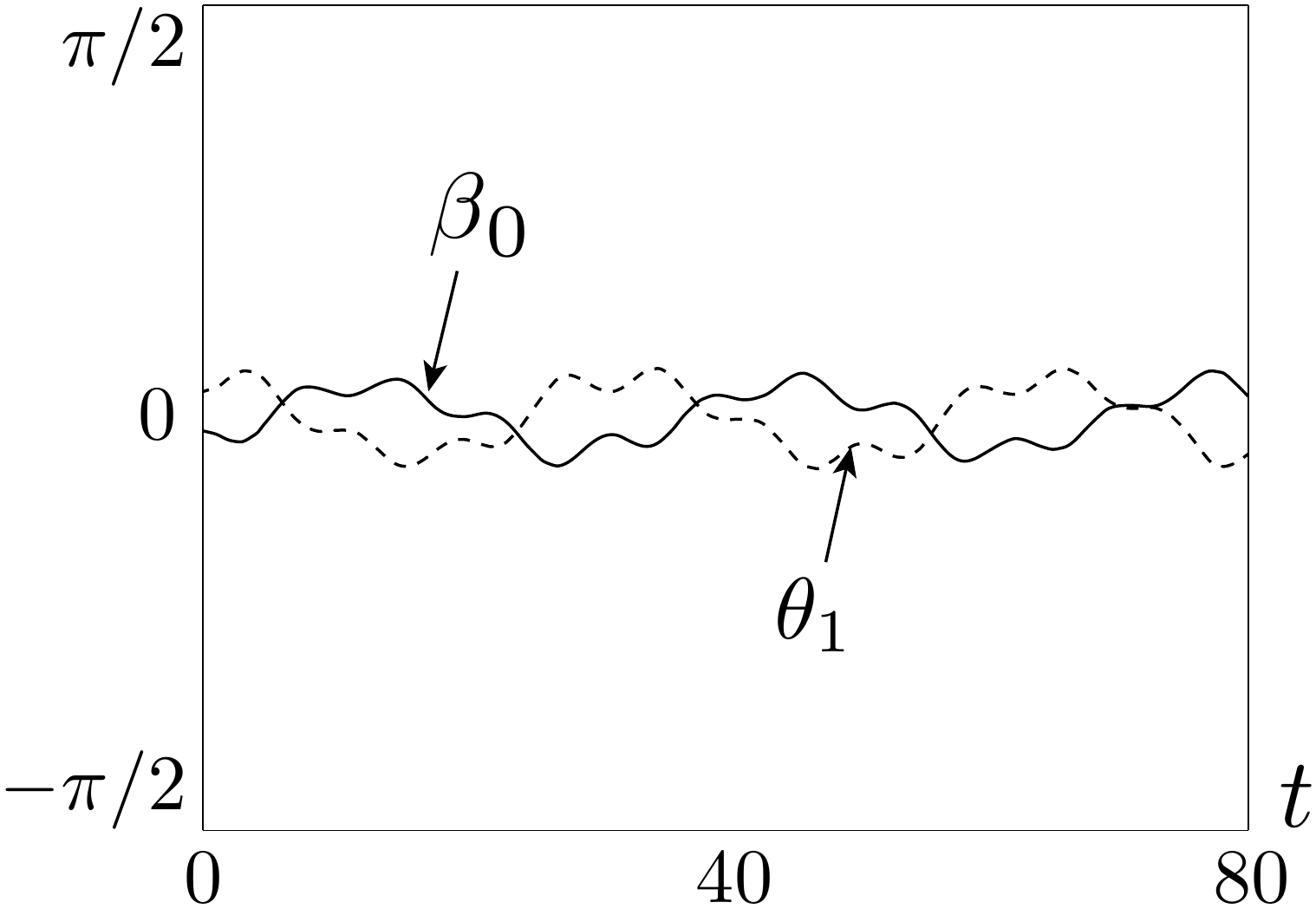} }
	\subfigure[$k = 0.5$]{
	\includegraphics[scale=0.28]{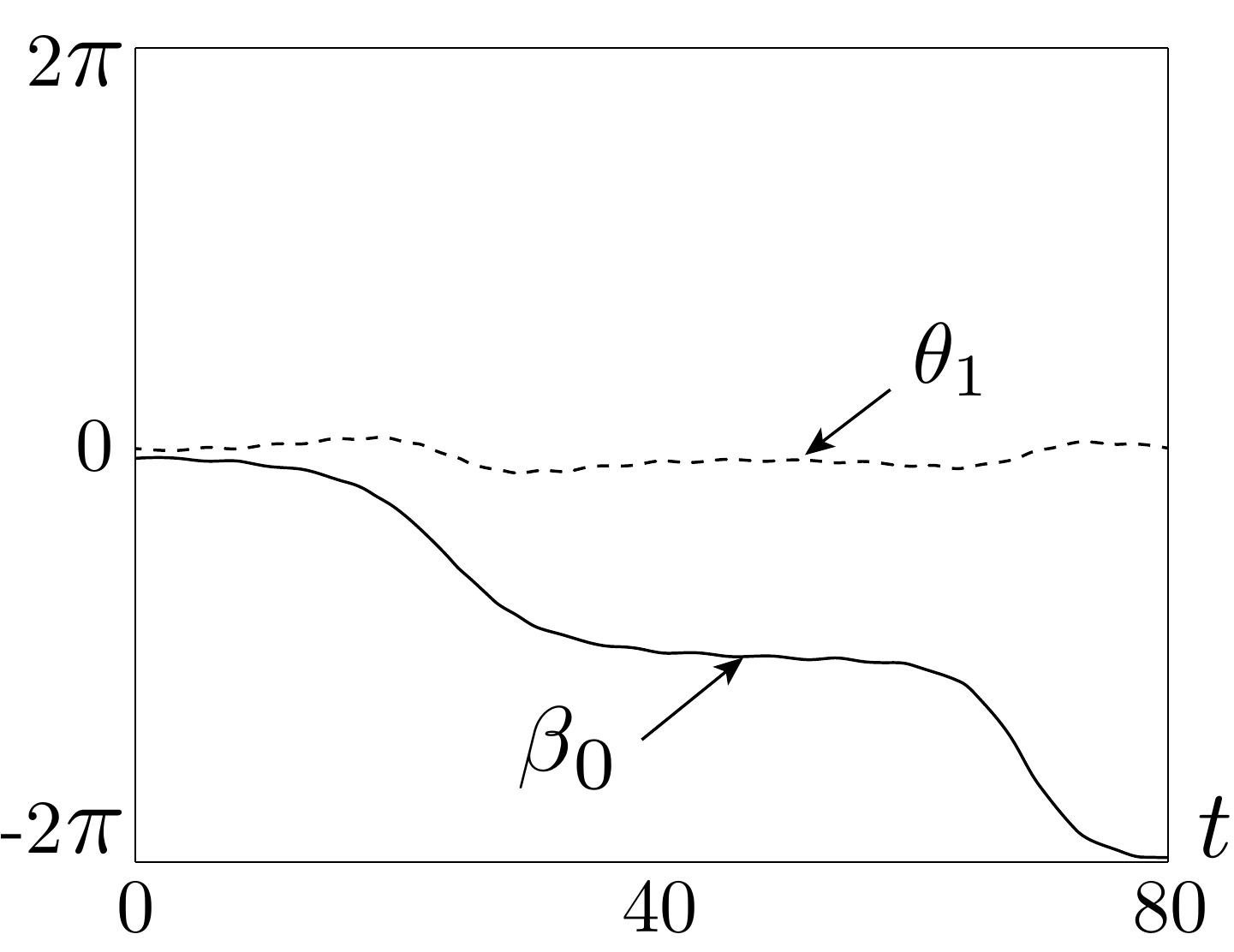} }
		\caption{\footnotesize Rotations $\beta_0$ and deformations $\theta_1$ versus time $t$ of two-link fish for $b=0.2$ and  (a) $k=0.2$, (b) $k=0.35$ and (c) $k=0.5$.   Initial condition is $\boldsymbol{\eta}_0 = [\, 1\,, 0\,, 0\,, 0.1\,, 0]^T$. Total integration time is 80.} \label{fig:angle}
	\end{center}
\end{figure}
\begin{figure}[!t]
	\begin{center}	
	\subfigure[$k = 0.2$]{
	\includegraphics[scale=0.285]{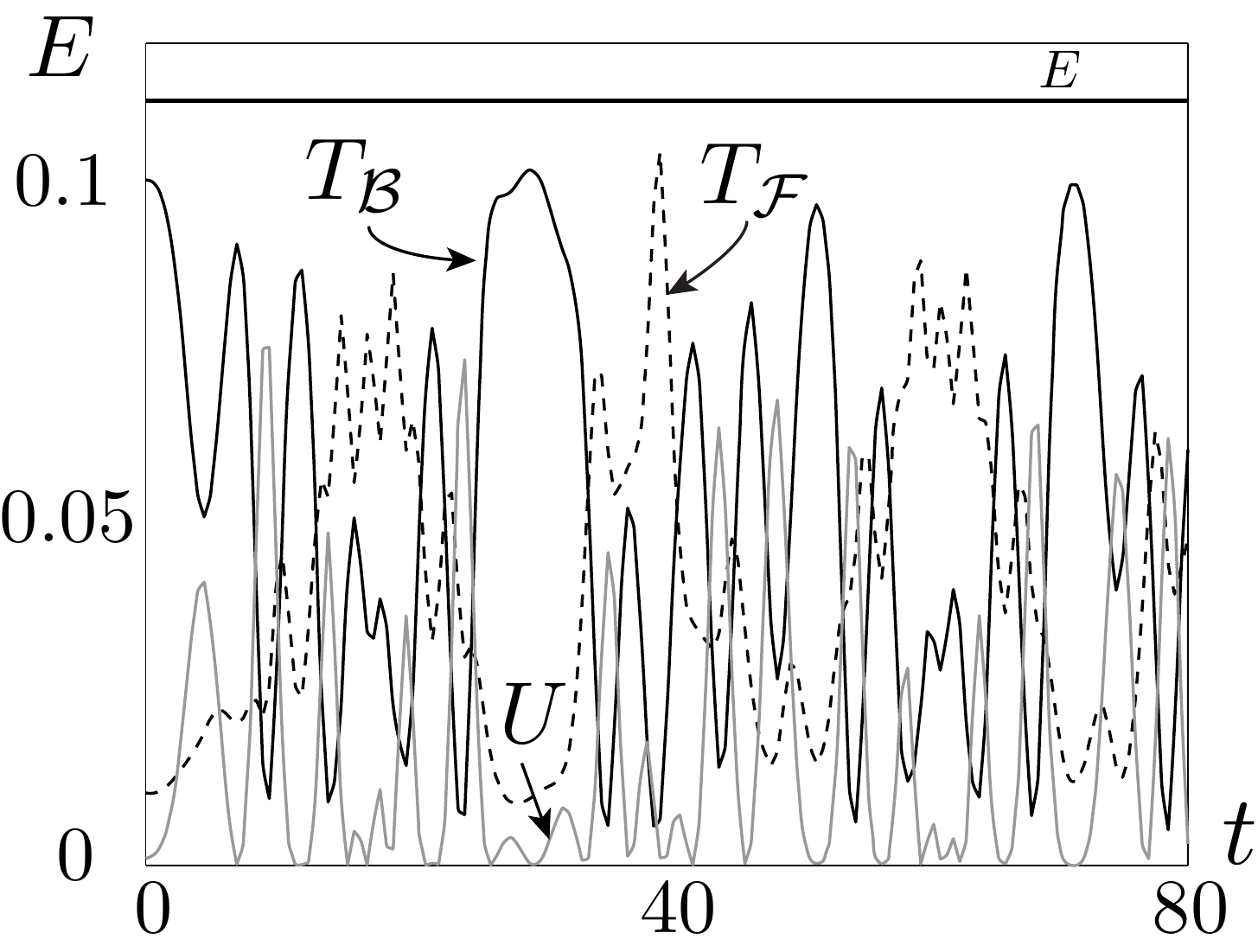} }
	\subfigure[$k = 0.35$]{
	\includegraphics[scale=0.285]{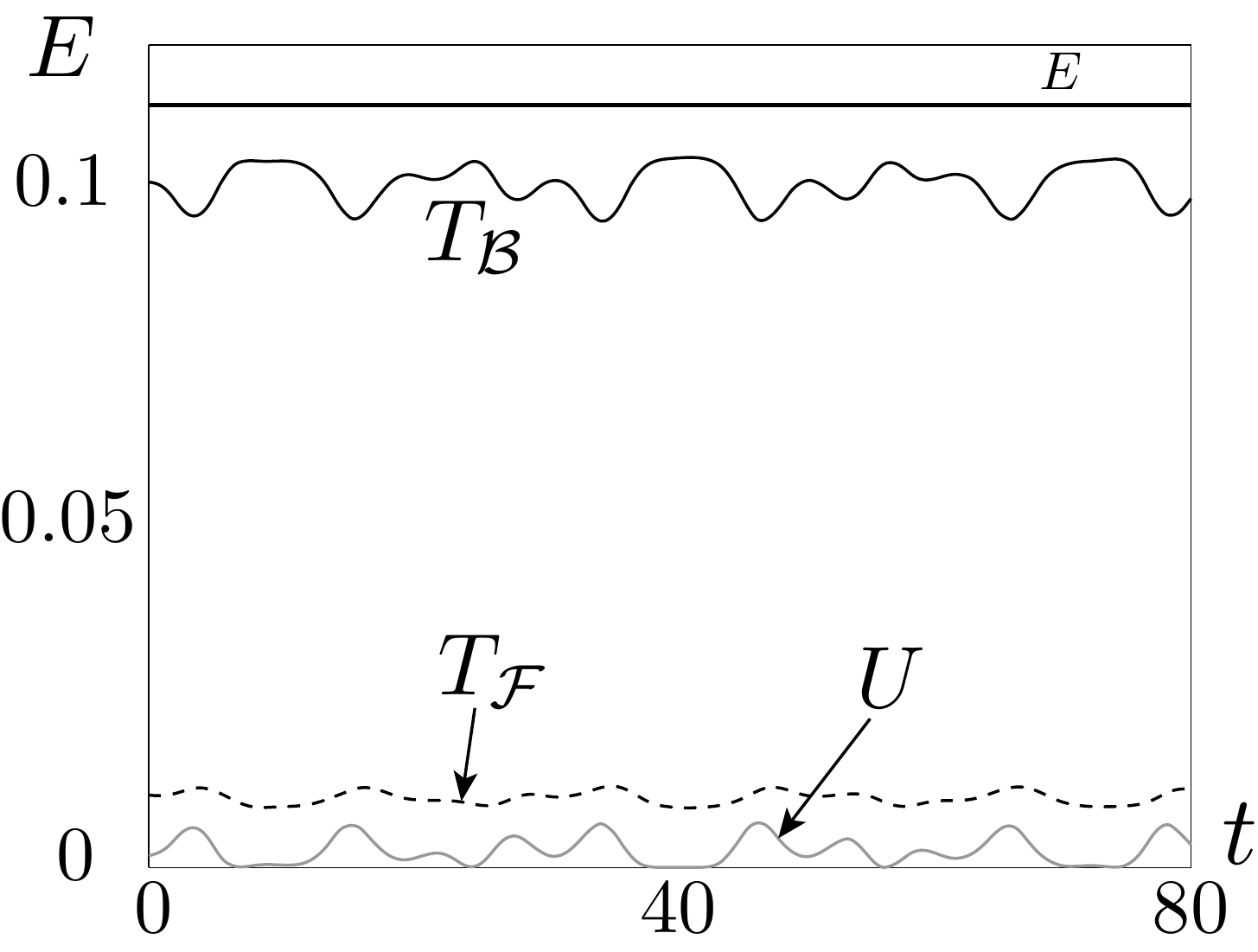} }
	\subfigure[$k = 0.5$]{
	\includegraphics[scale=0.285]{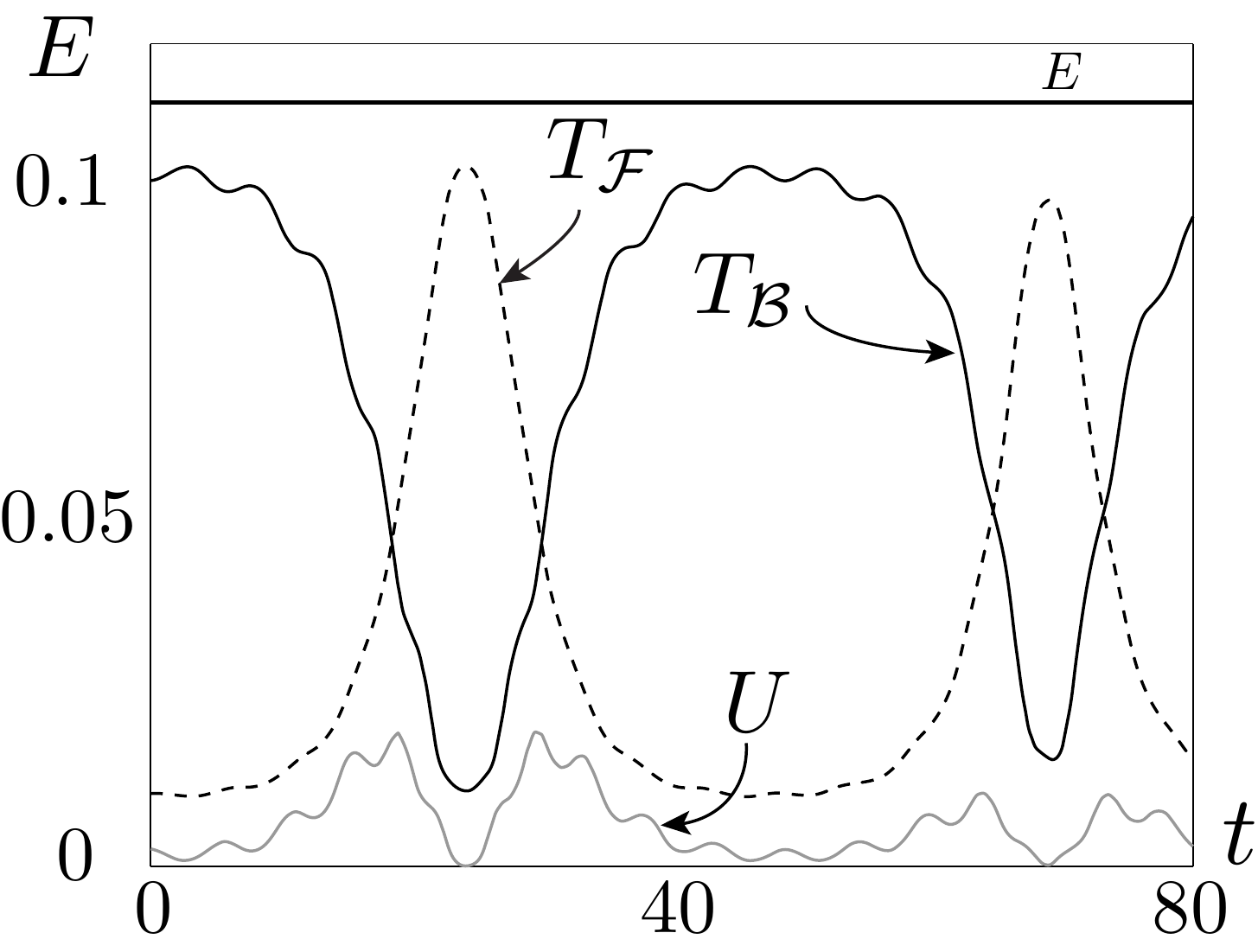} }
		\caption{\footnotesize Energy exchange of two-link fish for $b=0.2$ and (a) $k=0.2$, (b) $k=0.35$ and (c) $k=0.5$.   Initial condition is $\boldsymbol{\eta}_0 = [\, 1\,, 0\,, 0\,, 0.1\,, 0]^T$. Total integration time is 80.} \label{fig:energy}
	\end{center}
\end{figure}

Recall that for a single rigid ellipse (see figure~\ref{fig:equilibria}$(a)$), this relative equilibrium is unstable 
for all $b<1$. Due to the coupling between the translation and rotational motions in Kirchhoff's equations,
an initial perturbation causes an elliptic rigid body to tumble when moving along its elongated direction, \cite{Lamb1932,Leonard1997}. 
The  stability of this equilibrium is more interesting when the body is deformable.

We begin by examining the case of an articulated body made of two rigid ellipses (of non-dimensional minor axis $b$) connected via one hinge joint 
equipped with a torsional spring (of non-dimensional stiffness $k$).\footnote{we dropped the subscript on $k$ because there is only one 
spring} The relative equilibrium in non-dimensional form is $\boldsymbol{\eta}_e = [\, 1 \,, \,0 \,, \, 0 \,, 0 \,, \, 0\, ]^T$. We examine the passive 
behavior of the articulated body when we impose a small initial perturbation $\delta \bm{\eta}$ to this equilibrium, for instance, $\theta(0) = 0.1$. Figures~\ref{fig:trajectory} and~\ref{fig:angle} 
show the response of the nonlinear equations of motion~\eqref{eq:eom} for three different sets of parameter values, $b=0.1$ and $k=0.2$, $0.35$ and $0.5$, 
subject to initial conditions $\bm{\eta}_e + \delta\bm{\eta}$.
Figure~\ref{fig:energy} shows the exchange of energy for each of these cases. 
Given the non-dissipative nature of the model, the total energy $E=T_{\mathcal{B}} + T_{\mathcal{F}} + V$ is constant in time. But,  
there is exchange between the kinetic and potential energies of the system and consequently between 
the shape deformations and translational and rotational motions. 
For $k=0.35$, the motion of the articulated body (figure~\ref{fig:trajectory}$(b)$) remains close to the relative equilibrium and the body deformation $\theta_1$ and rotational motion $\beta_0$ (figure~\ref{fig:angle}$(b)$) remain bounded for the duration of this time integration. Indeed, figure~\ref{fig:energy}$(b)$ shows that the energy exchange is just right to set the body into deformation, that is to say, to set $\theta_1$ into oscillations, in response to the 
initial perturbation without causing the body to move far from the equilibrium solution. 
For the soft spring $k=0.2$, the articulated body moves away from the equilibrium trajectory (figure~\ref{fig:trajectory}$(a)$) while undergoing large deformations $\theta_1$ and large rotations $\beta_0$ (figure~\ref{fig:angle}$(a)$) which in turn are coupled to the translational motion.
This behavior indicates that the relative equilibrium is unstable for these parameter values. 
figure~\ref{fig:energy}$(a)$ shows that the deformation potential energy $V$ is comparable in value to the system's kinetic energy, which, by conservation
of total energy, indicates that the coupling between the shape deformations and the rotational and translational motion causes the system to move away from 
the equilibrium.
For the stiff spring $k=0.5$, the articulated body also moves away from the equilibrium  (figure~\ref{fig:trajectory}$(c)$) but in a way similar to the tumbling instability observed in the case of a single rigid body. Indeed, the body deformation $\theta_1$ is small in comparison to its rotational motion  (figure~\ref{fig:angle}$(c)$), and as in the case of a rigid body, it is the coupling between the rotational and translational motion that induces the instability, as evidenced from the energy exchange in figure~\ref{fig:energy}$(c)$. This is not surprising, for larger $k$, the articulated body is more rigid and its behavior approaches that of a rigid body.
Similar results are obtained for various values of the initial perturbations $\delta \bm{\eta}$ (results not shown).

We assess the linear stability of these trajectories. We let $\boldsymbol{\eta} = \boldsymbol{\eta}_e + \delta\boldsymbol{\eta}$, where $ \delta\boldsymbol{\eta}$ is small, substitute $\boldsymbol{\eta}$ into~\eqref{eq:eom} and neglect higher order terms. The linearized equations of motion can be expressed as
\begin{equation}
	\delta\dot{\boldsymbol{\eta}} = \mathbb{A}\, \delta\boldsymbol{\eta}\,,
	\label{eq:eom_linear}
\end{equation}
where the \emph{Jacobian} $\mathbb{A}$ is a $5\times 5$ matrix, and it has 5 eigenvalues $\lambda_\alpha, \alpha = 1,\ldots,5$. 
Given the conservative nature of the system (no energy dissipation), the corresponding relative equilibrium is said to be marginally stable if all eigenvalues of the 
linearized system have non-positive real parts. If at least one eigenvalue has positive real part, the equilibrium is linearly unstable. Remember that linear instability implies nonlinear instability but linear stability does not necessarily indicate nonlinear stability, see, for example,~\cite{Marsden1992}. 
\begin{figure}
	\begin{center}
	\includegraphics[scale=0.4]{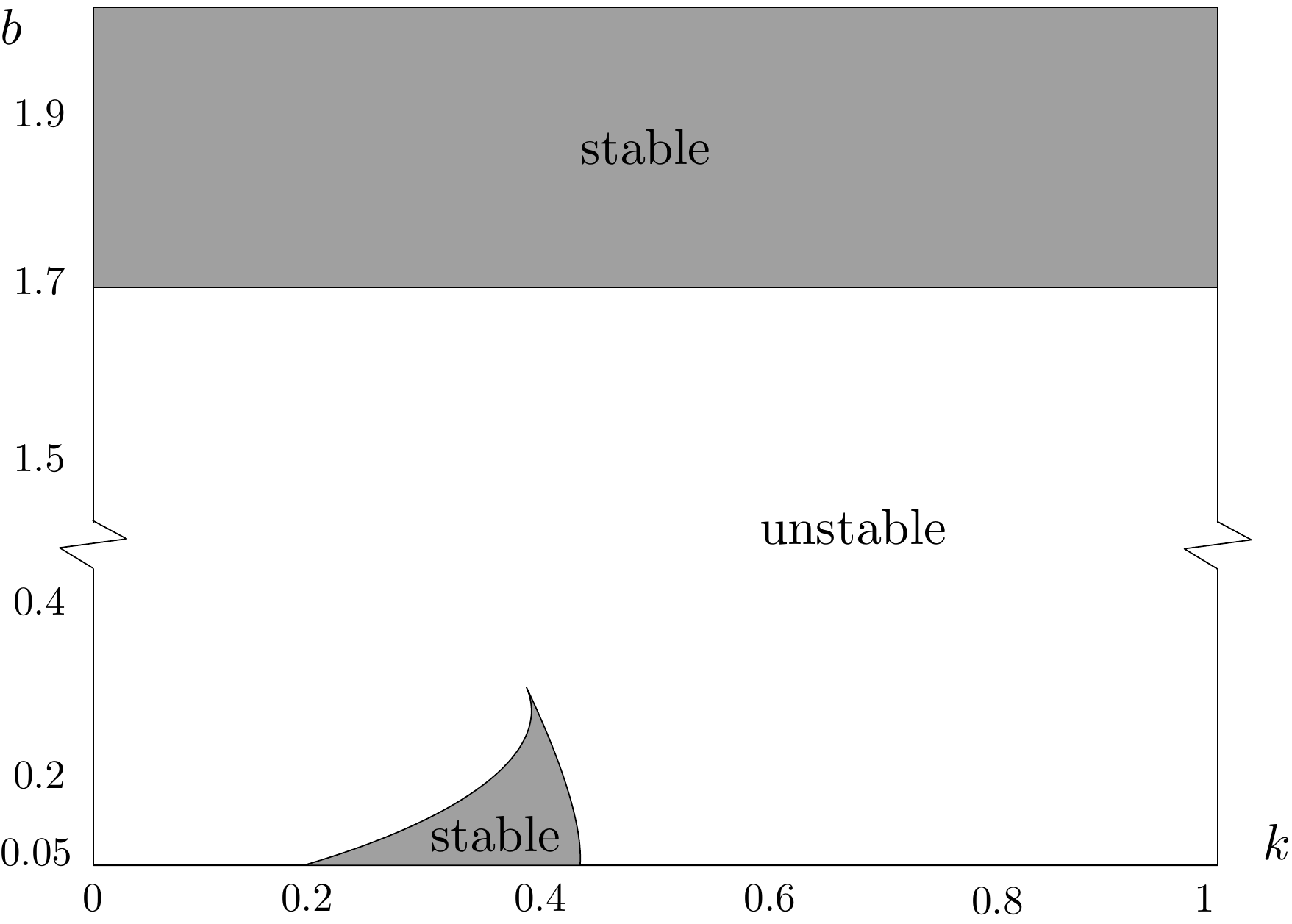}
		\caption{\footnotesize Parameter space $(b,k)$ of the two-link fish model. The regions for which the coast motion is linearly stable are highlighted in grey.} \label{fig:sr2}
	\end{center}
\end{figure}
\begin{figure}
	\begin{center}
		\includegraphics[scale=0.35]{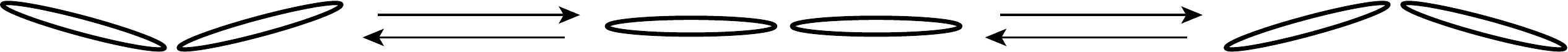}
		 \caption{\footnotesize Sketch of the dominant  mode of oscillation in the ``shark fin" stable region for the two-link model.}\label{fig:sm2}
	\end{center}
\end{figure}
For the two-link articulated body, we find that one of the eigenvalues is always 0, which reflects the symmetry associated with 
the fact that the equilibrium is invariant to translations in the direction of $\mathbf{b}_1$. The other eigenvalues depend on the two non-dimensional parameters 
of the system: the aspect ratio $b$ of the (identical) ellipses and the spring stiffness $k$. We conduct a parametric analysis of the stability as a function
of the two-dimensional parameter space $(b,k)$. Namely, we discretize a region of the parameter space from  $b \in [0.01,1]$ and $k\in [0, 2]$
using  increments of $\Delta b = 0.01$ and $\Delta k = 0.001$, and compute the eigenvalues of~\eqref{eq:eom_linear} for each point of the discrete parameter space to assess the stability of the system at that point. The results are shown in figure~\ref{fig:sr2}.  Two regions in the $(b, k)$ parameter space are found for which all eigenvalues have zero real part, which implies that the system is marginally stable in these regions. One region corresponds to $b \geq 1.68$ and the other one lies
in the area $b < 1$ and resembles a ``shark fin". The existence of the first stability region can be understood by analogy to a single ellipse. For a single ellipse, the coast motion is stable when $b \geq 1$, which means a rigid body is stable when it has circular shape ($b = 1$) or when it is translating along its semi-minor axis. For the two-link case, when $b \approx 1.68$, the articulated body becomes more circular in shape, granted not exactly circular, and  when $b > 1.68$, the two-link fish model is similar to a body moving along its minor-axis of symmetry. The ``shark fin" stability region is much more interesting. Indeed, in contrast to a single rigid body which is unstable for all $b < 1$, this stability region indicates that the coast motion can be passively stable when $b < 1$ by allowing the body to deform (via the deformation variable $\theta_1$) and by proper choice of body elasticity $k$. For a given aspect ratio, if $k$ is small (weak spring), the deformations are large
and the system is unstable. If, on the other hand, $k$ is too large (stiff spring), the deformations are small and the model is similar to that of a rigid body, which is also unstable. For an intermediate range of $k$ values which provide the appropriate bending resistance to perturbations, the system is stable.

It is informative to look at the linear modes of oscillation in the ``shark fin" stable region. Remember that one eigenvalue is zero, say $\lambda_1 = 0$. 
The remaining four eigenvalues are two pairs of pure imaginary complex conjugates, say, $\lambda_{2,3} = \pm\mu_1 i$ and $\lambda_{4,5} = \pm\mu_2 i$ with
$\mu_1 \leq \mu_2$. Let  $\zeta_{2,3} = \xi_1 \pm i \eta_1$ and $\zeta_{4,5} =\xi_2 \pm i \eta_2$ be the complex conjugate eigenvectors corresponding to $\lambda_{2,3}$ and $\lambda_{4,5}$, respectively.  $\mu_1$ and $\mu_2$ are the natural frequencies of oscillations and $\xi_1,\eta_1$ and $\xi_2,\eta_2$ are the associated linear modes of oscillations, with the smaller frequency $\mu_1$ dictating the dominant mode of oscillation. When $b = 0.1, k = 0.35$ for which the equilibrium is linearly stable, the smaller frequency of oscillation is $\mu_1 = 0.205$ and the associated mode of oscillation is given by $\xi_{1} = [0\,,0\,,-0.192\,,0.257\,,-0.214]^T$, $\eta_{1} = [0\,,0\,,0\,,-0.114\,,0]^T$. A sketch of this mode is depicted in figure~\ref{fig:sm2}. One can see that the two ellipses oscillate symmetrically about the hinge joint, that is to say, the dominant stable mode corresponds to a bending mode of deformation.



\begin{figure}
	\begin{center}
	\includegraphics[scale=0.7]{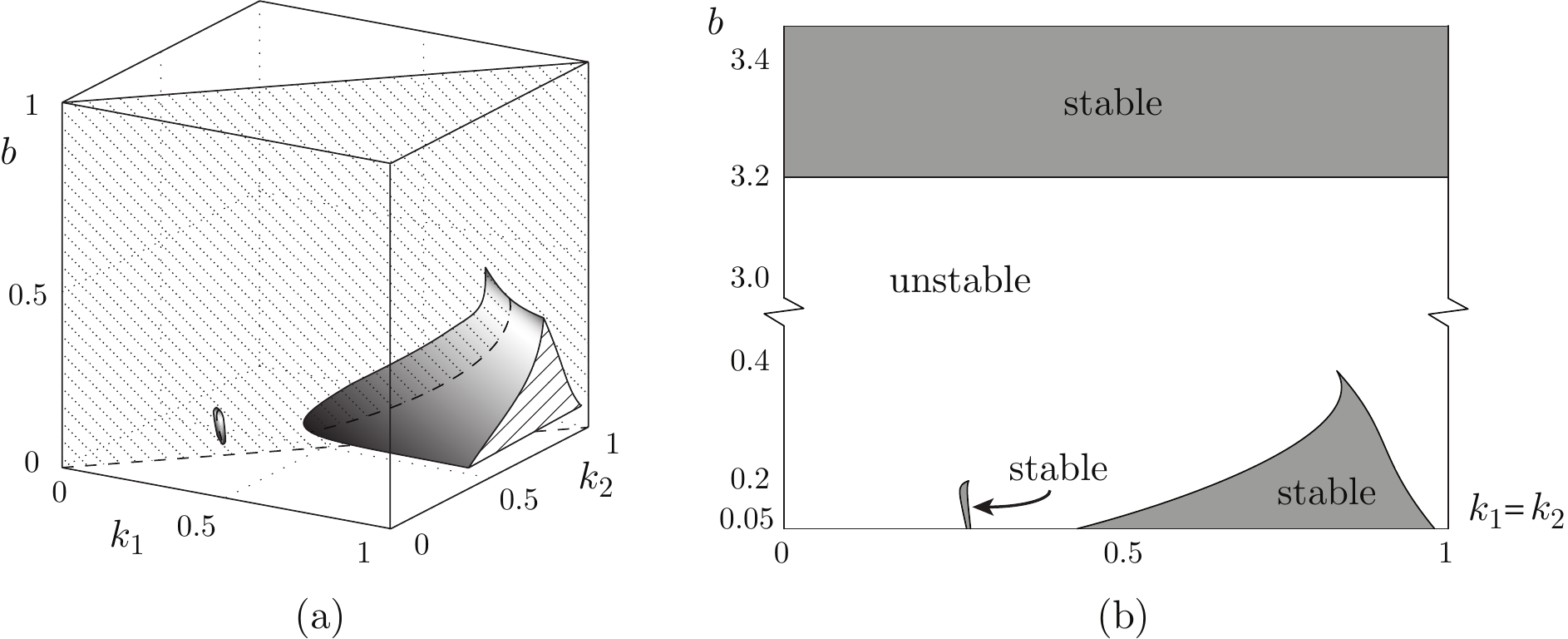}
	\caption{\footnotesize Three-link fish model:  (a) three-dimensional parameter space $(b, k_1, k_2)$. (b) planar cross section $(b, k_1 = k_2)$ Regions for which the coast motion is linearly stable are depicted in grey.} \label{fig:sr3}
	\end{center}
\end{figure}

\begin{figure}
	\begin{center}
		\subfigure[traveling-wave deformation]{
		\includegraphics[scale=0.3]{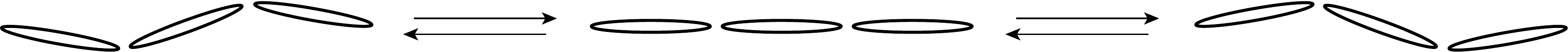}}
		\subfigure[bending deformation]{
		\includegraphics[scale=0.3]{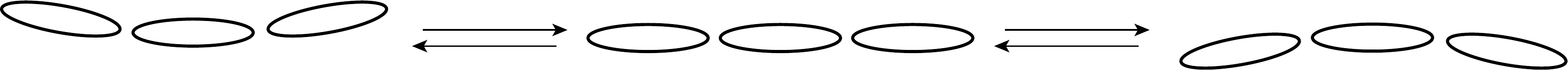}}
		 \caption{\footnotesize Sketch of the dominate modes of oscillation in the stable regions for the three-link fish model. (a) traveling-wave deformation mode is dominant in the small region. (b)  bending deformation mode is dominant  in the big region.}\label{fig:sm3}
	\end{center}
\end{figure}

We now ask how this stable region is affected when the articulated body has more deformation degrees of freedom. For concreteness, we consider an articulated body made of three identical links (of minor axis $b$) connected via two hinge joints equipped with torsional springs of stiffnesses $k_1$ and $k_2$. One now has two deformation variables $\theta_1$ and $\theta_2$ and  the independent variables are $\boldsymbol{\eta} = [\,u_0 \,, \, v_0 \,, \, \Omega_0 \,,\, \theta_1 \,,\, \dot{\theta}_1 \,,\, \theta_2 \,,\, \dot{\theta}_2\,]^T$. The parameter space is three-dimensional $(b,k_1,k_2)$. We follow the same procedure as above to assess the linear stability of the coast motion as a function of this parameter space.
Interestingly, one finds two stable regions in the area where $b<1$: a ``shark fin" region similar to the one obtained for the two-link model and a smaller region, see figure~\ref{fig:sr3}. To compare this stability result with the two-link case, we plot the cross section of the plane $k_1 = k_2$ in figure~\ref{fig:sr3}$(b)$. One can see that, in addition to the two stable regions when $b<1$, a third stable region exists for $b \geq 3.23$ which is analogous to the $b \geq 1.68$ region in the two-link model.  
The ``shark fin" region is also analogous to the stable region in the two-link model, though its area is larger than the latter. However, the small stable region does not have an analogy in the two-link model. To better understand this difference, we plot the dominant oscillation modes for both regions in $b < 1$ in figure~\ref{fig:sm3}. The dominant mode of oscillation in the ``shark fin" region is similar to that observed in the two-link model. Namely, the two side ellipses  are oscillating symmetrically relative to the middle ellipse which corresponds to a bending mode of deformation. Figure~\ref{fig:sm3}(a) illustrates the dominant mode in the smaller region. Here, the two side ellipses are oscillating anti-symmetrically relative to the middle ellipse, and the deformation mode corresponds to that of a traveling wave. This mode cannot exist in the two-link model. An interesting question which remains unanswered in this work is how these stability results get affected as the number of hinge joints goes to infinity, that is to say, as the body becomes completely deformable. We speculate that similar results, with both traveling-wave and bending modes of deformations, will hold depending on the elastic properties of the deformable body.
\section{Conclusions}\label{sec:discussion}

We used a deformable body consisting of an articulated body equipped with torsional springs at its hinge joints as a simple model 
to study the passive stability of the coast motion of an underwater swimmer. The coast motion, that is to say, the
motion of a swimmer with constant velocity along its elongated direction, is known to be intrinsically unstable if the body is
not deformable. Through a combination of numerical examples and linear stability analysis, we showed that 
the coast motion can be passively stabilized. We highlight the following main points:

\begin{enumerate}
\item[(i)] The surrounding fluid, even in an inviscid and irrotational context, couples the deformation of the body and its translational 
and rotational motion and allows for energy exchange between these motions in a way that causes initial perturbations to 
the coast motion to remain bounded.
\item[(ii)] Elasticity  of the body (finite $k$) when properly chosen can be the essential ingredient for the passive stabilization of an otherwise unstable coast motion.
\item[(iii)] The effects of viscous boundary conditions and the resulting generation of vorticity from the boundary layer region remain
an open question. 
One may think that viscosity and dissipation could enhance passive stability by damping out initial perturbations. 
However, it has been shown that the hydrodynamic forces created by vorticity shed from a bluff body
may cause vortex-induced vibrations (see, for example,~\cite{Griffin1984,WiGo2004}) which may in turn have destabilizing effects. 
These vortex-induced vibrations may be especially significant where the swimmer has a bluff shape 
($b/a>1.68$ for the two-link and $b/a>3.23$ for the three-link). Their effects on the motion of the more fusiform-shaped swimmer
($b/a<1$) is less clear. On the one hand, the deformation modes may add to the bluffness of the swimmer's shape. On the other hand, 
separation and shedding may be hindered by a traveling-wave deformation if the wave speed is faster than the swimming speed; see, for example,
Borazjani \& Sotiropoulos (2008), (2009a,b);~\cite{ShZh2003, TaTo1974}. 
In reality, drag forces will slow down the passively swimming fish and passive locomotion
cannot be maintained. The fish will have to actively flap in order to overcome the drag forces and continue swimming -- thus the burst phase in the burst and coast cycle. During the burst phase, the fish could potentially correct any destabilizing forces due to vortex shedding.
\item[(iv) ]While we restricted the discussion in this paper to planar motions, the main stabilizing mechanisms discussed here are also present 
in a three-dimensional environment but for only a subset of the 3D perturbations, namely, for planar perturbations 
(see~\cite{Jing2011}). 
These passive mechanisms are not sufficient to stabilize out-of-plane perturbations and one needs to incorporate
active control mechanisms. The next steps in the development of this class of models are to address
three-dimensional motion planning, implement  control strategies for both stabilization and produce complex maneuvers, such as 
turning and fleeing responses, as well as to examine multi-body synchronization (see, for example,~\cite{NaKa2007,TcCrLe2010}) 
and its role in facilitating and/or stabilizing the coast motion.
\end{enumerate}

\paragraph{Acknowledgments.} This work is partially supported by the National Science Foundation through the CAREER award 
CMMI 06-44925 and the Grant CCF 08-11480.

%
%
\bibliographystyle{jfm}

\end{document}